\documentclass[sigconf]{acmart}

\usepackage[utf8]{inputenc}

\usepackage{soul}
\usepackage{xcolor}
\usepackage{xspace}
\usepackage{url}
\usepackage{graphicx}
\usepackage{listings,multicol}
\usepackage[caption=false]{subfig}  
\usepackage[export]{adjustbox}
\usepackage{amsmath}
\usepackage{textcomp}
\usepackage{breakurl} 
\usepackage{listings}
\usepackage{cleveref}
\usepackage{algorithm}
\usepackage[noend]{algpseudocode}
\usepackage{balance}

\newcommand{\ie}{\emph{i.e.,}\xspace}
\newcommand{\eg}{\emph{e.g.,}\xspace}

\newcommand{\toolname}{\textsc{Lupa}\xspace}

\begin{document}

\title[\toolname: A Framework for Large Scale Analysis of the Programming Language Usage]{\toolname: A Framework for Large Scale \\ Analysis of the Programming Language Usage}

\author{Anna Vlasova}
\affiliation{
  \institution{\textit{JetBrains Research}}
}
\email{anna.vlasova@jetbrains.com}

\author{Maria Tigina}
\affiliation{
  \institution{\textit{JetBrains Research}}
  \institution{\textit{ITMO University}}
}
\email{maria.tigina@jetbrains.com}

\author{Ilya Vlasov}
\affiliation{
  \institution{\textit{Saint Petersburg State University}}
}
\email{ilyavlasov2011@gmail.com}

\author{Anastasiia Birillo}
\affiliation{
  \institution{\textit{JetBrains Research}}
}
\email{anastasia.birillo@jetbrains.com}

\author{Yaroslav Golubev}
\affiliation{
  \institution{\textit{JetBrains Research}}
}
\email{yaroslav.golubev@jetbrains.com}

\author{Timofey Bryksin}
\affiliation{
  \institution{\textit{JetBrains Research}}
}
\email{timofey.bryksin@jetbrains.com}

\begin{abstract}
    In this paper, we present \toolname~--- a platform for large-scale analysis of the programming language usage. 
    \toolname is a command line tool that uses the power of the IntelliJ Platform under the hood, which gives it access to powerful static analysis tools used in modern IDEs. 
    The tool supports custom \textit{analyzers} that process the rich concrete syntax tree of the code and can calculate its various features: the presence of entities, their dependencies, definition-usage chains, etc. 
    Currently, \toolname supports analyzing Python and Kotlin, but can be extended to other languages supported by IntelliJ-based IDEs. 
    We explain the internals of the tool, show how it can be extended and customized, and describe an example analysis that we carried out with its help: analyzing the syntax of ranges in Kotlin.
\end{abstract}

\maketitle

\section{Introduction}\label{sec:introduction}

Software engineering becomes an ever more important part of our life and enters almost all of the existing industries and fields of study~\cite{boehm2006view}.
To answer the emerging challenges and to incorporate the earned experience, new programming languages continue to be developed, and the existing ones are evolving~\cite{flauzino2018you, karus2011study, mateus2020adoption}.
This process can take many forms: new libraries that implement novel techniques~\cite{zhauniarovich2015stadyna, li2014self}, new language features~\cite{dyer2013large}, or even new programming languages altogether~\cite{bose2018comparative, mateus2019empirical}.

To help with this process, it is important to keep track of these features, their changes, and their prevalence~\cite{dyer2014mining, peng2021empirical, li2021empirical}.
Narrower topics can also be investigated, such as configuration settings~\cite{schermann2018structured}, popularity trends~\cite{borges2016understanding}, popular testing practices~\cite{kochhar2013adoption}, etc.

Several approaches exist to study language features on a large scale. 
Among these, two main groups can be highlighted. 
The first one~\cite{sokol2013metricminer, trautsch2020smartshark, spadini2018pydriller, maj2021codedj} is focused on the meta-information about projects, like the information about the number of stars or collaborators on GitHub. 
These tools usually do not have access to the code of the projects and therefore can not be used to study programming languages.
The second group of tools can take an existing dataset and perform the analysis of the code itself~\cite{dyer2013boa, peng2021empirical, li2021empirical, guilardi2020androidproptracker}. 
This is important in a large number of cases, since the code contains a rich set of information about how language constructs are used. 
Unfortunately, such platforms mostly focus on just several prominent languages and features within them, do not use industry-level tools to parse and analyze the source code, and instead opt for implementing their own specific pipelines. For this reason, while such tools are in principle extendable, it is often difficult to actually extend them in practice. 

To overcome the described problems of the code-based tools, in this paper, we present \toolname~--- an extendable framework for analyzing fine-grained language usage built on top of the IntelliJ Platform~\cite{kurbatova2021intellij}.
\toolname is a command line tool that uses the power of the IntelliJ Platform~\cite{intellij} under the hood to perform code analysis using the same industry-level tools that are employed in IntelliJ-based IDEs~\cite{kurbatova2021intellij}, such as IntelliJ IDEA~\cite{intellijIDEA}, PyCharm~\cite{pycharm}, or CLion~\cite{clion}.
This platform provides a rich concrete syntax tree called Program Structure Interface (PSI)~\cite{psi}, and an API for its traversal (visitors, name resolving, finding usages, etc.). 
Currently, our framework supports the analysis of two languages: Python --- a mature language most popular in the data science and machine learning domains~\cite{githubAnnualReport}, and Kotlin --- a relatively young but quickly growing language~\cite{kotlinstat}.

The most important feature of \toolname is the ease of its extension. 
The current implementation is written in Kotlin, but both Java and Kotlin can be used to extend the tool.
To perform a custom analysis within the supported language, one needs only to add a new \textit{analyzer}: the necessary filters expressed in terms of PSI nodes. 
For example, to detect all usages of a certain function, one needs to visit all the nodes that correspond to function calls and filter them by name. 
\toolname will then run this analysis on all projects in the provided dataset and output the results. 
Our framework can also be extended to any language that PSI supports (this includes Java, JavaScript, PHP, C/C++, Go, and many others)~\cite{kurbatova2021intellij}.

To demonstrate the usefulness of \toolname and to provide an example of its usage, we describe a study of range declarations in Kotlin. 
In this analysis, we collected a dataset of 10 thousand Kotlin projects and wrote several analyzers for \toolname that studied different syntactic ways of declaring ranges (\textit{e.g.}, \texttt{X..Y}, \texttt{X.rangeTo(Y)}, etc.) and the context of their use.
To get the context of the usage, \toolname traversed the syntax tree of the code to find the necessary parent node (for example, to see that the range is used within an \texttt{if} condition).
The results show that different kinds of declarations are used unevenly and in different contexts. This information was used by the development team of Kotlin to plan further experiments with the syntax of ranges.

\toolname is available for researchers and practitioners on GitHub: \url{https://github.com/JetBrains-Research/Lupa}. We believe that the framework can be of use for researchers for a wide array of studies on source code, as well as for the developers and maintainers of libraries that are interested in how their features are used.
\section{Background}\label{sec:background}

A lot of existing research has been dedicated to studying large corpora of code~\cite{dyer2014mining, chatziasimidis2015data, peng2021empirical, li2021empirical, schermann2018structured, borges2016understanding, kochhar2013adoption}, which involved developing tools, frameworks, and platforms for their processing~\cite{sokol2013metricminer, trautsch2020smartshark, spadini2018pydriller, maj2021codedj, dyer2013boa, peng2021empirical, li2021empirical}. 
In this section, we list the most notable ones.

The first category of tools allows calculating various metrics without code analysis, for example, the number of stars for a GitHub repository. 
The tools in this category can be implemented as online platforms such as \textsc{MetricMiner}~\cite{sokol2013metricminer} and \textsc{SmartSHARK}~\cite{trautsch2020smartshark}, or as desktop applications such as \textsc{PyDriller}~\cite{spadini2018pydriller} and \textsc{CodeDJ}~\cite{maj2021codedj}.

Online platforms from this category~\cite{sokol2013metricminer, trautsch2020smartshark} are limited to the datasets that are loaded into their database.
First of all, this drawback limits the scope of analysis to languages present in the dataset, usually only popular ones. 
In addition, it is impossible to view the data in dynamics, since the database contains a specific revision of the data (which can also be outdated).
At the same time, such platforms allow performing analytical queries and filtering repositories without much effort, since the user needs to simply open a website, write a query, and get the results.

Desktop tools~\cite{spadini2018pydriller, maj2021codedj} are more difficult to use as they require a laptop or a server with the suitable characteristics, free hard disk space, and the effort required to install the tool.
Such tools usually work with GitHub API and allow not only running some queries but also downloading repositories that satisfy these requests.
These tools can be adapted to new queries by extending filter engines and can be used with specific languages and datasets.
The main disadvantage of such platforms is that the analysis in this case is limited only to studying various types of meta-information about the project, the authors, etc., which makes it impossible to study source code in detail.

The second distinct group of tools usually analyze pre-gathered datasets or have simple GitHub scrapers, and allow scanning code files, even performing static analysis.
One such popular tool is \textsc{Boa}~\cite{dyer2013boa} --- a language and an infrastructure for analyzing large-scale datasets. 
This language allows building an abstract syntax tree (AST) for code files and filtering projects by a simple parsing of this tree.
The tool can traverse the AST and filter its nodes by text conditions, \textit{e.g.}, if a node has \textit{null} in its name.
\textsc{Boa} is actively maintained~\cite{boaAbout} and popular among researchers~\cite{boasCited}, but it does not have the ability to use other static analysis features, such as searching for references or resolving variable types.

Another prominent tool is \textsc{PyScan}~\cite{peng2021empirical}, which allows scanning Python repositories, resolving all dependencies and types. 
The tool defines 22 kinds of popular language constructs for Python applications and analyzes their usage.
The approach is based on building an AST, designing patterns for each language feature of interest, and using an external tool for dynamic type inference.
Such a tool is difficult to extend even inside Python, since to support new patterns, it requires one to design the given pattern, as well as, if necessary, implement additional tools for static code analysis.

In another work, Li developed a tool~\cite{li2021empirical} aiming at building an extended AST for Bash projects and analyzing popular constructs of the language.
The author uses a set of Python scripts to download and filter GitHub data, and the IntelliJ Platform~\cite{intellij} to build the tree and carry out static analysis.
However, the tool is distributed as a replication package with no possibilities for extension to other languages and analyzers.

It can be seen that the described solutions generally only support prominent languages (usually just one or several each), and also do not use industry-level parsing capabilities, instead implementing their own analysis tools (different ones for different languages), which makes it more difficult to extend them to custom needs. 
To overcome these drawbacks, we developed \toolname.
\section{\toolname}\label{sec:internals}

\begin{figure}[h!]
\centering
    \includegraphics[width=0.9\columnwidth]{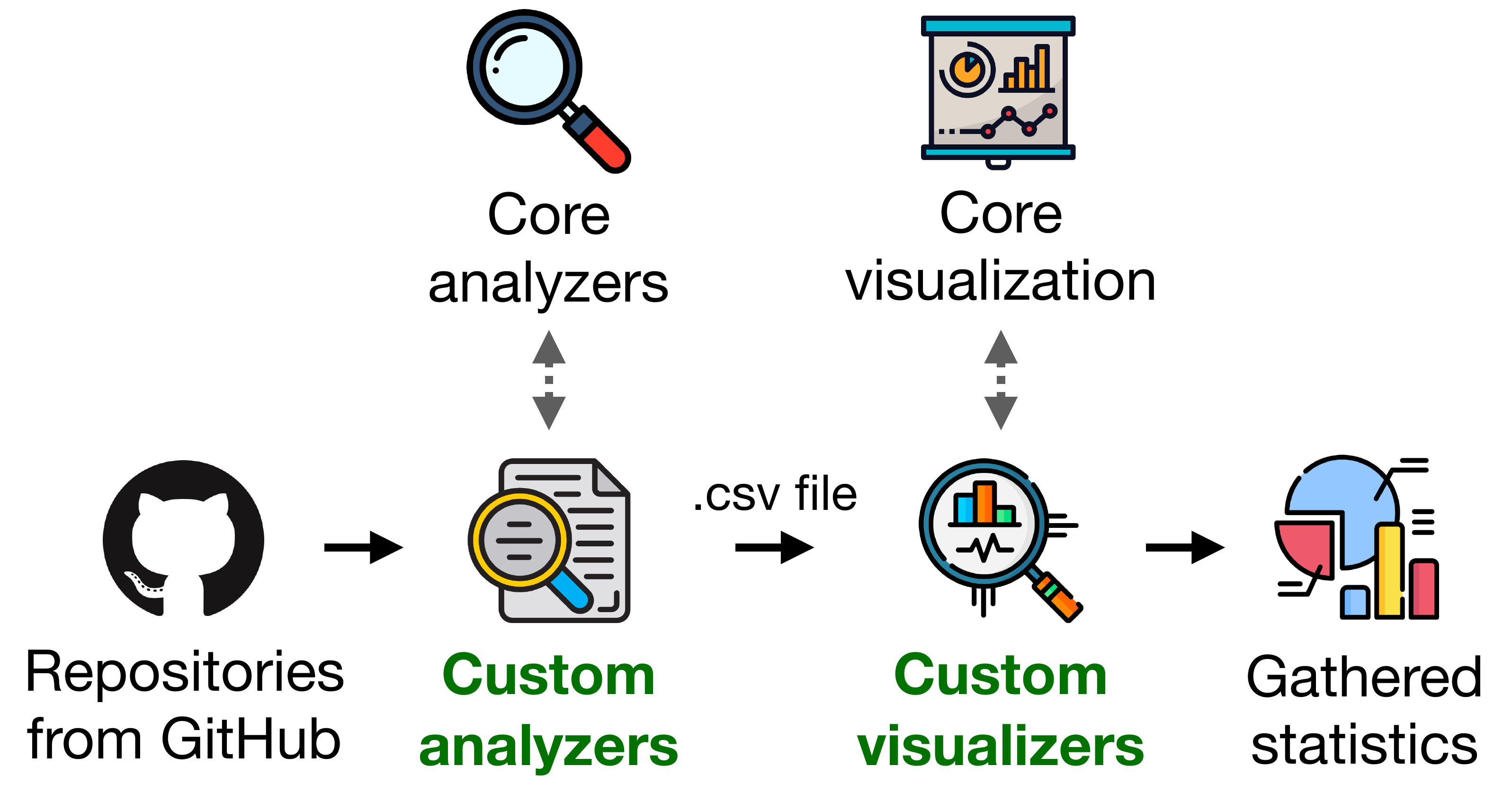}
    \centering
    \caption{An operating pipeline of \toolname.}
    \label{fig:pipeline}
    \vspace{-0.5cm}
\end{figure}

\toolname is a platform for large-scale analysis of the programming language usage.
Specifically, \toolname is implemented as a plugin for the IntelliJ Platform~\cite{intellij} that reuses its API to launch the IDE in the background (without graphical user interface, GUI) and run the necessary analysis on every project in the given dataset.
The workflow of \toolname is demonstrated in Figure~\ref{fig:pipeline}.
The user needs only to write a set of instructions about which nodes of the syntax tree they want to analyze, and launch the tool.
In this section, we explain the architecture and the tooling behind \toolname, describe the process of extending it to custom analysis and visualization, provide several example use cases, and discuss the existing limitations.

\vspace{-0.1cm}

\subsection{Internals}

The tool is written in Koltin and operates as a plugin for the IntelliJ Platform.
Apart from traditional plugins, when the plugin operates in the opened IDE via GUI, the IntelliJ Platform also supports the so called \textit{headless mode}.
In this mode, the plugin operates as a command line program that launches the IDE in the background, without GUI, and then the IDE performs the necessary work. 

This operation pipeline is extremely convenient for our task for three main reasons. 
Firstly, the plugin can launch different IDEs, meaning that it can support different languages. 
Secondly, this allows us to use the full power of industry-level code analysis on large datasets, like finding usages of functions, without having to open each project manually. 
Thirdly, this allows us to use the existing tooling without parsing or processing the code ourselves. 
Mainly, it gives us access to a particularly useful part of the IntelliJ Platform --- Program Structure Interface (PSI). 
The PSI tree is a rich concrete syntax tree that allows us to perform efficient and deep static analysis on the opened project.
Unlike a simple abstract syntax tree, PSI contains information about the semantics of code, can index the entire project to connect entities between different files, resolve dependencies, build control flow graphs, etc.

\toolname operates as follows. 
To perform the analysis, the tool needs two obvious components: a \textit{dataset} and \textit{analyzers}, \ie sets of instructions of what PSI tree nodes need to be analyzed and how. 
The tool takes the list of links to GitHub projects as input, checks them for duplicates using GitHub API, and then downloads the dataset.
The analyzers are implemented using the tool's infrastructure (see~\Cref{extensions} for details), several analyzers can be run together in a single passing of the dataset.
When the arguments are passed, the plugin can be run with one simple CLI command. 
It is also possible to launch the tool in the batch mode.
In this case, the tool will process the given number of projects (the size of the batch that can be configured), save all results individually, and then merge them in the end.
This can be used to save the results iteratively as checkpoints in case of external problems when processing a large dataset.
\toolname launches the IDE in the background, which then opens the projects in the batch one by one. 
This happens the same way as if one would open the project in an IDE --- the necessary dependencies are automatically downloaded, indexed, and resolved, the semantic and syntactic PSI models of the project are built. 

When the project is opened, the plugin gains access to the PSI tree of each necessary file. 
Then it simply executes instructions from the analyzers and saves their results into a CSV file. 
When this is done for all the files, the results are saved, the project is closed, and \toolname moves to the next project in the batch. 
The final results can be analyzed further or visualized: the plugin has several core functions for visualization, \textit{e.g.,} plotting interactive bar charts for visualizing CSV data that can be customized by the users via changing the parameters of the chart or adding custom functionality, and then used in Jupyter notebooks.
\vspace{-0.4cm}
\subsection{Extensions}\label{extensions}

\begin{figure}[t]
\centering
    \includegraphics[width=\columnwidth]{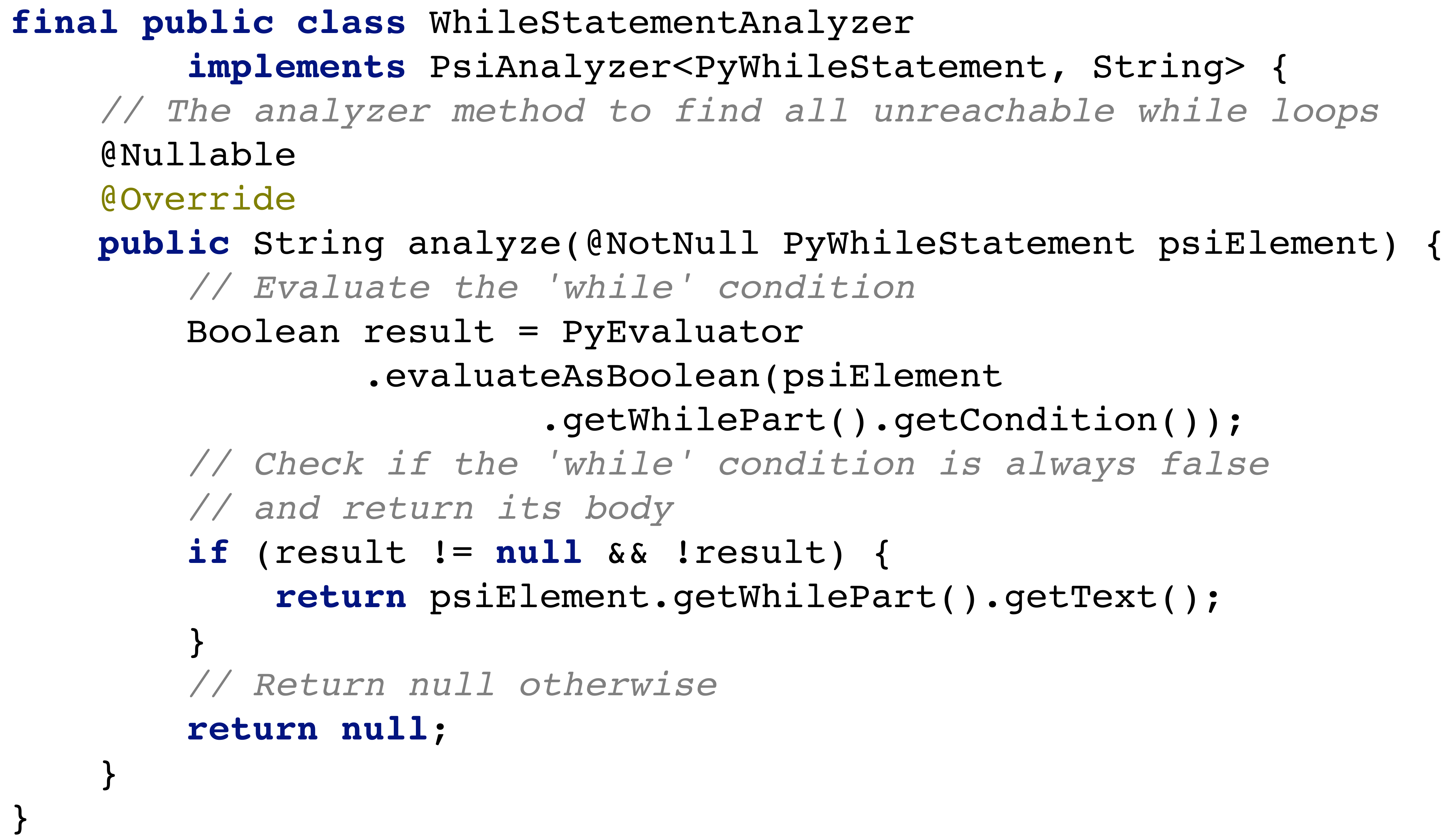}
    \centering
    \caption{An example of a simple analyzer written in Java. It finds all the \texttt{while} loops, evaluates their conditions, and collects their content if the condition is always false.}
    \label{fig:analyzer}
    \vspace{-0.4cm}
\end{figure}

The main feature of \toolname is that one can write the necessary analyzers themselves in a pretty straightforward way. 
Since the plugin is written in Kotlin, this can be done in either Kotlin or Java.
Specifically, each analyzer is a class that describes a filter that should be applied to the PSI tree to obtain the necessary result. 
This can be as simple as ``filter only numerals'' or something more complex like ``find all the unreachable \texttt{while} loops and save their content''. 
While the second example sounds complicated, in reality, it is merely a series of filters and checks applied to PSI nodes. 
The implementation of this particular 
analyzer for Python code is presented in Figure~\ref{fig:analyzer}. 
The analyzer is written in Java, but can also be written in Kotlin.

Besides accessing different PSI elements, Figure~\ref{fig:analyzer} highlights the use of the \texttt{evaluateAsBoolean} method (a part of the IntelliJ Platform) to check the boolean value of a statement. So while something trivial as \texttt{while False} can be found using regular expressions, \toolname would be able to find something like \texttt{while 2+2 != 4} as well, which simpler tools would not be able to find. 
This example analyzer demonstrates the power of PSI for code analysis.

The tool provides a flexible architecture that supports different types of analyzers: simple filtering by node types, filtering by context (the use of a given node in the scope of another node, \eg using the \texttt{print} function within the \texttt{try} branch), and others. 
\toolname is fully documented and contains a dozen examples of analyzers that we used for our purposes. 
Also, importantly, several analyzers can be run together in a single passing of the dataset.

Finally, the analyzers can be written not only for Python and Kotlin, but also for any other language that PSI supports. Since the IntelliJ Platform is used as the base for IDEs for different languages (JavaScript, TypeScript, PHP, Go, and others), all these major languages can be analyzed. Doing this requires one to access the necessary PSI elements for the chosen language and add appropriate dependencies to the necessary parts of the IntelliJ Platform.

\subsection{Use Cases}

\toolname can be helpful in several different scenarios. 
Firstly, it can be used to perform general large-scale studies about how various languages are used~\cite{dyer2014mining, peng2021empirical, li2021empirical}. 
This can be as simple as counting the usage of specific keywords and constructs~\cite{peng2021empirical, li2021empirical}, or something more complex that involves dependencies between different entities, referencing certain fields, etc.~\cite{dyer2014mining, li2021empirical}. 
This can be used to track the evolution of a language or to compare them between each other.

In particular, tracking changes in a language can be useful in education~\cite{parker2006formal, canou2017scaling, odiete2017recommending}. 
Languages often develop quite quickly, new libraries and frameworks appear~\cite{endres2005survey, wang2019various}, and at the same time, new courses emerge for learning new technologies~\cite{sharov2021analysis}. 
In this case, \toolname can help teachers in choosing technologies to study and finding examples of their usage.

Another way the framework can be useful is to track the usage of a specific library~\cite{kula2018developers, Landman2017ChallengesFS}. 
While the simple fact of its use can be studied using, for example, Code Search on GitHub~\cite{githubSearchCode}, \toolname allows implementing more complex analyzers to see in which context specific functions are used, what parameters they receive, what types they work with, etc. 
This can be used by researchers in general or by developers of a specific library themselves~\cite{kula2018developers}.
\vspace{-0.1cm}

\subsection{Limitations}

\toolname has several important limitations that should be mentioned. 
The main one is the operating speed of the tool. 
Deep code analysis requires a longer time, since in order to process usages and connect entities, it is necessary to index the project and resolve the dependencies. 
\toolname can thus be most useful on datasets of moderate size.
In our studies, we used a dataset of 10 thousand Kotlin projects from GitHub, running the tool on a laptop with the following characteristics: 2.4 GHz 8-Core Intel Core i9 processor and 32 GB of RAM. 
Running \toolname on this dataset took on average 6.5 seconds per project for Kotlin.
This time measurement does not take into account downloading all the necessary libraries, since this needs to be done only once.

The second limitation is that while \toolname is extendable, its scope is still limited by only the languages that are supported by PSI. 
The list of these languages is rather long, but it does not include some specific languages, \textit{e.g.}, OCaml.
Finally, while writing a new analyzer is simple enough, it is still more difficult than writing a script with a regular expression or an SQL query. 
Writing the necessary analyzer in Java or Kotlin can be viewed as using a DSL of sorts, however, PSI and the IntelliJ Platform are well-documented.

Even though these limitations exist, we believe that they do not invalidate the usefulness of the tool, and that it can be of use for researchers and practitioners in many cases.
\section{Example Analysis of Kotlin Ranges}\label{sec:analysis}

In our research group, we employed \toolname for a number of different studies, both practical and theoretical. 
In this section, we describe one of these studies that showcases the usefulness of the tool as well as the potential diversity of its application.

\textbf{Background.} 
Kotlin supports several different ways to write numerical ranges. 
There are three main options to define an ascending range (\texttt{X..Y}, \texttt{X until Y}, and \texttt{X.rangeTo(Y)}) and one main way for descending ranges (\texttt{X.downTo(Y)}). 
Different options are more convenient in different situations, and they also differ by their rules of inclusion and exclusion of the borders. 
For the purposes of language evolution, our colleagues from the Kotlin development team wanted to see which of them are more popular and where.

\textbf{Research questions.}
With this research, we wanted to answer two questions.
\textbf{(1)} Which ways of declaring ranges are popular? 
\textbf{(2)}~Are some ways more popular than others when used in a specific context (loops, conditions, etc.)?

\textbf{Dataset.}
To collect the data, we used the Web service developed by Dabic et al.~\cite{Dabic:msr2021data}. 
This tool provided us with a list of all GitHub projects that have Kotlin as the main language, are not forks of other projects, and have at least 10 stars. 
Using the tool, we downloaded the projects in April of 2021, which resulted in 10,536 projects.
We deleted all duplicates, \textit{e.g.}, repositories that were moved, and the final dataset contained 10,442 projects.

\textbf{Methodology.}
To answer the research questions, we created two different analyzers for \toolname that were then run simultaneously.
Both analyzers processed only Kotlin files, and firstly detected all the necessary PSI nodes (functions that correspond to the four types of range declarations). 
To answer the first question, these nodes were simply counted. 
To answer the second question, the second analyzer then traversed the parents of each node until it found a node from the list of language constructs that were of interest to the Kotlin development team: \texttt{for} loops, \texttt{while} loops, \texttt{if} conditions, etc. 
This allowed us to see where the range is located in the program's body. 
Finally, we wrote some visualizations that built the necessary charts.
To do this, we extended the core visualization functions from our framework.

\textbf{Results.}
\textit{Popularity.} Our analysis discovered more than 70 thousand usages of ranges in the studied projects. 
Two of the four ways of declaration are significantly more popular than the other two: \texttt{X..Y} covers 52.2\% of cases, \texttt{X until Y} covers 45.6\% of cases, while the other two syntactic forms only cover the remaining 2.2\%.
This result can help the Kotlin development team focus their attention on specific types of ranges in future language improvements.

\textit{Context.} We discovered that the studied declarations are used very differently in various contexts. 
While in general, as mentioned above, \texttt{X..Y} and \texttt{X until Y} are more or less similar in popularity, they are used within different language constructs.
For example, the former is more than three times as popular when used in \texttt{if} conditions, while the latter is two times as popular when used in \texttt{for} loops.
Such specific distribution is very important for language developers, since it allows them to develop custom recommendations in IDEs that depend on the context.
An example would be different code completion recommendations in different code contexts.
It is important to note that it is impossible to obtain accurate results like these using simple regular expressions or full text search.
\section{Conclusion}\label{sec:conclusion}

In this paper, we presented \toolname, a framework for large scale analysis of the programming language usage.
\toolname is a plugin for the IntelliJ Platform that launches the IDE in the background, which then performs the necessary analysis.
This allows \toolname to reuse a lot of the platform's industry-level abilities and features: rich concrete syntax tree, syntactic and semantic models, etc.
This helps our framework to overcome the drawbacks of many existing tools, since it is integrated into a well-established and documented platform and can be easily extended.
To extend the tool for custom analysis, a researcher only needs to write the necessary analyzer, \textit{i.e.}, a class that details the filters that should be applied to the syntax tree of the code. 
The tool currently supports the analysis of Kotlin and Python, but can be extended to other languages as well.

To showcase \toolname, we developed a dozen analyzers and described an example study.
In it, we used the tool to study different ways of defining ranges in Kotlin along with the context where they are used.
The results demonstrated the diversity among the usage of ranges in Kotlin, and highlighted the usefulness of our framework for such research.

In the future, we plan to extend \toolname to support more languages, expand the list of pre-configured example analyzers, and develop more custom visualizations to support the full pipeline of a large scale analysis.

\balance

\bibliographystyle{ACM-Reference-Format}
\bibliography{paper}

\end{document}